\documentclass[prd,twocolumn,a4paper,superscriptaddress,floatfix]{revtex4}
\usepackage{graphicx}
\begin{document}
\newcommand{\be}{\begin{equation}}
\newcommand{\ee}{\end{equation}}

\title{Cosmic Strings} 
\author{A. Ach\'ucarro} 
\affiliation{Institute Lorentz of Theoretical
  Physics, University of Leiden, 2333CA Leiden, The Netherlands}
\affiliation{Department of Theoretical Physics, University of the
  Basque Country UPV-EHU, 48940 Leioa, Spain} 
\author{C.J.A.P.  Martins} 
\affiliation{Centro de Astrof\'{\i}sica, Universidade do
  Porto, Rua das Estrelas s/n, 4150-762 Porto, Portugal}
\affiliation{DAMTP, CMS, University of
  Cambridge, Wilberforce Road, Cambridge CB3 0WA, United Kingdom}
\date{1 February 2008}
\begin{abstract}
Draft version of the contribution to  {\it Encyclopaedia of Complexity and Systems Science}, R. Myers (Ed.), Springer New York (2009)
\end{abstract}
\maketitle

\section{Definition of the Subject and its Importance}

Cosmic strings are linear concentrations of energy that form whenever
phase transitions in the early universe break axial symmetries \cite{Vilenkin}, as
originally shown by Kibble \cite{Kibble76,Kibble80}. They are the result of
frustrated order in the quantum fields responsible for elementary
particles and their interactions.  For about two decades, motivation
for their study was provided by the possibility that they could be
behind the density inhomogeneities that led to the observed
large-scale structures in the universe. Precision observations,
particularly of the cosmic microwave background radiation,
have limited strings to a sub-dominant role in structure formation.
Instead, the inhomogeneities appear
to be consistent with a period of cosmological inflation, but it turns
out that particle-physics models of the early universe that predict a
period of inflation very often also predict the generation of cosmic
strings at the end of it \cite{Jeannerot97,Jeannerot03}.

More recently, interest has been revived with the realization that
there may be strong links between field theory cosmic strings and
fundamental strings. The latter are the supposed ultimate building
blocks of matter, and in their original context of superstring theory
were thought to be microscopic. However, in its modern
version---sometimes referred to as M-theory---it is possible and
perhaps even mandatory to have macroscopic (cosmological-sized)
fundamental strings \cite{Majumdar02,Sarangi02,Davis05}. Their behavior is expected to
be quite similar to that of field theory cosmic strings, although
there are some important differences so they may in principle be
observationally distinguishable. Being relics of the phase transitions
that produced them, cosmic strings provide us with a unique window
into the early universe. If they are stable and survive for a
significant amount of time (possibly even up to the present day), they
may leave an imprint in many astrophysical and cosmological
observables, and provide us with information on fundamental physics
and the very early universe that would otherwise be inaccessible to
us. On the other hand, gaining a quantitative understanding of their
properties, interactions, evolution and consequences represents a
significant challenge because of their intrinsic complexity. Their
non-linearity is particularly noteworthy, with highly non-trivial
feedback mechanisms between large (cosmological) and small
(microscopic) scales affecting the network dynamics.  Considerable
reliance, therefore, must be placed on numerical simulations, which
are technically difficult and computationally costly. A complementary
approach is the use of analytic or semi-analytic models, usually to
describe the large-scale features of the networks.

The basic picture of the cosmological evolution of string networks
that has emerged for the simplest (Goto-Nambu) networks is of a
scaling solution with about 40 long strings always stretching across
each horizon volume plus a population of loops (other string types can
lead to a different behaviour).  It is then possible to estimate their
cosmological implications quantitatively.  For example, these strings
continuously source
gravitational perturbations on sub-horizon scales. The one parameter
in these models is the energy scale of the phase transition at which
the strings are created.  The astrophysical consequences of strings
stem from the non-trivial gravitational field around a string \cite{Vilenkin81}. Particles in the vicinity of a static
straight string feel no gravitational acceleration, because in general
relativity tension is a negative source of gravity and, since tension
equals energy per unit length, their effects cancel. The space-time
around the string is locally, but not globally, flat. In fact the
space is conical, with a deficit angle \be
\alpha=8\pi\frac{G\mu}{c^4}\, , \ee where $\mu$ is the energy per unit
length; the simple way to picture this is to imagine a plane in which
an angular wedge $\alpha$ has been removed and the edges glued
together.  For cosmologically interesting strings the deficit angle
ranges from a few seconds of arc to a few millionths of a second of
arc.

\section{Introduction}

To understand the cosmological evolution
and effects of cosmic strings we start in this section with a quick
summary of basic cosmology concepts that will be needed later and a
description of the simplest type of cosmic strings in which the main
features are already apparent.  But, first, a warning about units:
from now on we will set the speed of light to unity $c=1$, so we
measure distances in light-travel time, and masses in units of Energy
(and viceversa, using $E = mc^2$).  Boltzmann's constant is set to
unity, so temperature is measured in units of mass/energy (using $E =
K_B T$). Finally, we set Planck's constant to unity, $\hbar = 1$, and
measure all lengths and masses in units of Planck's length and
mass/energy:  $l_P = 1.62 \times 10^{-35}$ m , $M_P = 2.18 \times
10^{-8} $kg $= 1.22 \times 10^{19}$ GeV/c$^2$.  In these units,
Newton's constant is given by $G =  M_P^{-2}$. 

The early universe is very smooth. To a very good
approximation it is a homogeneous and isotropic spacetime described by
a single variable: the rate of expansion of its three-dimensional
spatial sections. In Einstein's general relativity this spacetime is
described by the flat Friedmann-Robertson-Walker (FRW) metric
\begin{equation} ds^2 = dt^2 - a^2(t) [\vec{dx} \cdot \vec{dx}] \ \ , 
\label{FRWt}\end{equation} where $\vec x$
are fixed (comoving) spatial coordinates and $a(t)$ is the scale
factor that determines the fractional or \emph{Hubble expansion rate}
\begin{equation} H(t) = {1 \over a(t) } {da \over dt} \ \ .
\label{defH}\end{equation} The time coordinate $t$ is known as
cosmological time; to analyse cosmic string evolution we will also
need a different time parametrization known as conformal time, $\tau$.
They are related by $ d\tau = {dt / a(t)}$, leading to the metric
\begin{equation} ds^2 = a^2(\tau) [d\tau^2 - \vec{dx} \cdot \vec{dx}]\ \ . 
\label{FRWtau}
\end{equation}

The age of the universe is currently estimated to be about 13.7 billion years \cite{Spergel07}.
The universe starts very hot and dense and is
cooled by the expansion, with the temperature decreasing as $T(t) \sim
a(t)^{-1}$.  The Hubble expansion rate is determined by the energy
contents of the universe. In a universe dominated by radiation or very
relativistic matter (the hottest, earliest stages), the scale factor
evolves as $a(t) \sim t^{1/2}$ and the energy density in radiation as
$\rho_{\rm radiation} \sim a(t)^{-4} \sim t^{-2}$.  The energy density
of non-relativistic matter is inversely proportional to volume
$\rho_{\rm matter} \sim a(t)^{-3}$
and eventually takes over (after about 4000 years), leading to a
period of matter domination, during which $a(t)= t^{2/3}$ and
therefore $\rho_{\rm matter}\sim t^{-2}$.  More recently --about five
billion years ago-- we have entered an epoch of accelerated expansion
due possibly to a cosmological constant or some unknown form of
\emph{dark energy} whose energy density is constant in time
$\rho_{\rm dark \ energy} \sim \ $const. Dark energy
should not be confused with dark matter, an unknown form of matter
whose presence we can detect through its gravitational effects but
that does not interact with electromagnetic fields and so in
particular does not emit light --hence the adjective ``dark''--. In
the currently accepted cosmological model the energy density in the
universe today would be dominated by dark energy (about 74\%),
followed by about 22\% dark matter and only about 4\% of regular
(baryonic) matter \cite{Spergel07}.
Dark matter is widely believed to be a particle still to be discovered.

The universe today is far from smooth, but the structure we observe on
the scale of clusters of galaxies is consistent with the gravitational
collapse of tiny primordial density inhomogeneities $\delta \rho /
\rho \sim 10^{-5}$ at the time the cosmic microwave background (CMB)
radiation was emitted. The CMB is the oldest radiation we observe,
dating back to the time when the universe was only 380000 years old.
At this epoch the primordial plasma cooled enough to allow
the formation of the first atoms (a process known as recombination),
and it became transparent to photons (which is
referred to as decoupling). Before that moment, the photons behave
like a fluid that is strongly coupled to the protons and electrons. An
overdense region in the baryon fluid would like to contract but the
photon pressure pushes it back, causing both fluids to oscillate.
These oscillations are imprinted in the cosmic microwave background
and can be detected today in the form of Doppler peaks in its power
spectrum.

The spectrum of density inhomogeneities has been accurately measured
in the CMB and found to be near scale-invariant and of the right
magnitude to produce the structure we observe. The perturbations to
the FRW metric can be classified as \emph{scalar} (overall changes to
the Newtonian gravitational potential), \emph{vector} (associated with
velocity and/or rotational effects) and \emph{tensor} (transverse
traceless perturbations to the spatial metric, such as
gravitational waves). Each of these affects the CMB in different
ways, so their relative contributions can in principle be
observationally distinguished.  Finally, Thomson scattering of the
anisotropic distribution of the CMB photons is particularly important
during decoupling and recombination, and induces a partial linear
polarization of the scattered radiation, at a level that is around ten
percent of the anisotropy. Detection of this polarization signal is at
the borderline of the sensitivity of ongoing experiments at the time
of writing, but is expected to become standard with forthcoming
experiments.

The energy density of a network of cosmic strings in the linear
scaling regime is $\rho_{\rm strings} \sim t^{-2}$ and therefore it
remains a constant fraction of the dominant form of energy during
matter or radiation domination. Numerical estimates for the simplest,
Goto-Nambu, networks suggest the fraction is around $100 G\mu$. Provided
the string mass is not close to the Planck scale, this is small enough not
to disturb the cosmological evolution; at the same time, for a broad
range of values of $G\mu$ this is large enough to be detectable in precision
experiments today.  Other string types (see section VI) may have
larger or smaller fractions or qualitatively different signatures. In
particular, networks that do not reach linear scaling may come to
dominate the energy density (which rules them out) or to disappear
completely.

The simplest field theory model that produces cosmic strings has a single
complex scalar field $\Phi$ (this is shorthand for a function $\Phi(t,
{\vec x})$ with complex values that do not change under coordinate
transformations).
Let us assume that the Hamiltonian determining the field dynamics is
invariant under an axial symmetry such as a phase rotation,
$\Phi\to\Phi e^{i\theta}$.  For example, take the potential energy \be
\int d^3x V= \int d^3 x
\frac{\lambda}{2}\left(|\Phi|^2-\eta^2\right)^2 \ee where $\lambda$ is
a dimensionless coupling constant and $\eta$ is an energy scale
related to the temperature of the symmetry breaking transition. This
has a set of degenerate ground states: the minimum of the potential in
field space is the circle $|\Phi| = \eta$, known as the \emph{vacuum
manifold}. Any configuration $\Phi(t, {\vec x}) = {\textrm const.} = \eta e^{i
\chi }$ with $\chi$ real and constant is a possible ground state or
\emph{vacuum}, irrespective of the value of the phase $\chi$.

\begin{figure}
\includegraphics[width=3.5in,keepaspectratio]{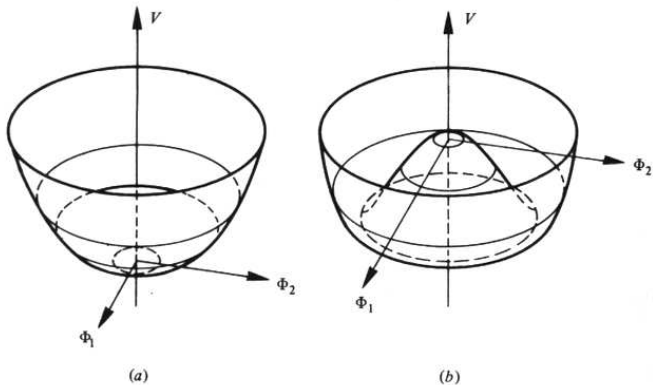}
\caption{\label{figpot}The effective potential energy $V$ for a simple
string-forming field theory model. The (a) and (b) plots correspond to
the high and low temperature configurations, respectively. For simplicity
the complex field $\Phi$ has been split into two real scalar fields, $\Phi_1$
and $\Phi_2$.}
\end{figure}

Figure \ref{figpot} illustrates what happens.
At high temperature the field fluctuations are large enough to make
the central peak around $|\Phi| = 0$ irrelevant, and the effective
potential is symmetric and has a minimum there. As the temperature
falls the energy will eventually be too low to permit fluctuations
over the peak, at which point the field will tend to settle towards
one of the ground states.  The random choice of minimum in this
condensation process then breaks the original axial
symmetry. This is the case, for instance, in superfluid $^4$He.

When a large system goes through a phase transition like this, each
part of it has to make this random choice, which need not be the same
everywhere. The minimization of gradient terms in the energy of the
system tends to make it evolve towards increasingly more uniform
configurations, but causality (the principle that no information can
travel faster than light) imposes that this evolution can only happen
at a limited rate. As a result one expects many domains,
each with an uncorrelated choice of ground state. Where these domains
meet there is some probability of forming linear defects---cosmic
strings---around which the phase angle varies by $2\pi$ (or possibly
multiples thereof). This is the Kibble mechanism. Notice that the
field vanishes at the string's core, so
there is trapped potential energy (as well as gradient energy). These
strings are known as \emph{global} strings because the axial symmetry
that is broken below the phase transition is ``global'',
that is, the transformation $\Phi\to\Phi e^{i\theta}$ is independent
of position.

The next step is to consider charged scalar fields interacting with an
electromagnetic field. The best known example of a
symmetry--breaking transition of this kind is the condensation of
Cooper pairs in a superconductor, that has the effect of making
photons massive below the critical temperature (in this case the axial
symmetry is of the ``local'' of ``gauge'' type).  The cosmic strings
that result are magnetic flux tubes that do not dissipate because the
magnetic field is massive outside the string core.

This type of vortex was first discussed by Abrikosov \cite{Abrikosov57} in the context of type II
superconductors. Nielsen and Olesen \cite{Nielsen73} generalized these ideas to the relativistic quantum field theory
models used in particle physics, in particular the Abelian Higgs model
which is a relativistic version of the Landau-Ginzburg model of
superconductivity, governed  by the action.

\begin{equation}
S =\!\! \int d^4x\left[|\partial_\mu \Phi - iq A_\mu \Phi|^2 - \frac1
4{F_{\mu\nu} F^{\mu\nu}} 
- {\frac\lambda 2}(|\Phi|^2 - \eta^2)^2 \right] \ \ .
\end{equation}
$A_\mu$ is the gauge field and $\Phi$ is a complex scalar of charge
$q$ ($q=2e$ in superconductors, where $\Phi$ is the Cooper pair
wavefunction). The second term is the usual Maxwell action for the
electromagnetic field, $F_{\mu\nu} = \partial_\mu A_\nu - \partial_\nu
A_\mu$. The energy per unit length of a straight, static string lying on
the $z$-axis is
\begin{equation}
E =\!\! \int d^2x\left[|\partial_x \Phi - iqA_x \Phi|^2 + |\partial_y\Phi - 
iqA_y\Phi|^2 + \frac1 2 B^2 +{\frac\lambda 2}(|\Phi|^2 - \eta^2)^2 \right]
\end{equation}
where $B = \partial_x A_y - \partial_y A_x$ is the $z$-component of
the magnetic field. Finite energy configurations must have $|\Phi| =
\eta$ (the {\it vacuum manifold} is still a circle) but the phase of
$\Phi$ is undetermined provided the gradient terms and the magnetic
field go to zero fast enough. This condition allows for finite energy
solutions $A_t = A_r = A_z = 0$, $\Phi (r, \theta) \sim \eta e^{i n
  \theta}$, $ A_\theta (r, \theta) \sim n / (qr) $, as $r \to \infty$,
in which the total magnetic flux in the plane perpendicular to the
string is quantized,
\[ \int d^2 x B = \oint {\vec A}\cdot {\vec dl} =  {2\pi n \over q}
\]
$n$ is the winding number of the string. 
If the constants $\lambda$ and $q$ are such that
fluctuations in the scalar field $\Phi$ and the gauge field $A_\mu$
have equal masses, it is possible to show that the string saturates an
inequality of the form
\[\rm Energy \ per \ unit\
length \geq \ constant \ x \ | magnetic \ flux| \] known as the
Bogomolnyi bound \cite{Bogomolnyi76}. In this case, parallel strings at close
range exert no force on each other and there are
static multivortex solutions \cite{Taubes81}.  If the mass of the scalar excitations is lower (higher) than
that of the gauge excitations, parallel strings will attract (repel).

More complicated particle physics
models --in particular those describing the early universe-- involve
gauge symmetries that generalize the electromagnetic interaction,
mediated by photons, to more complicated interactions such as the
electroweak or Grand Unified interactions.  The messenger fields that
play the role of the photons may be massless in the early universe and
become massive following a symmetry-breaking transition, and cosmic
strings carry the magnetic flux of these other massive gauge fields
(not the electromagnetic field).

From a cosmological point of view, the gauge field has the important
effect of making the gradient terms decay exponentially fast
away from the string so the energy per unit length of these
strings is finite.  Abrikosov--Nielsen--Olesen strings have no
long-range interactions, so their evolution is dominated by their
tension and is well described in the thin string or Goto-Nambu
approximation.

Field continuity implies that a string of this kind cannot simply come
to an end: it must form a closed loop or extend to infinity,
and it cannot break into segments. For this reason, strings,
once formed, are hard to eliminate.  In the absence of energy loss
mechanisms, the strings would eventually dominate the energy density
of the universe. On the other hand, the strings can decay into
radiation, they may cross and exchange partners, and they may
also cross themselves, forming a closed loop which may shrink and
eventually disappear. The outcome of these competing mechanisms is
that the network is expected to reach a scale-invariant (or
\emph{scaling}) regime, where the network's characteristic length
scale is proportional to the size of the horizon. We will discuss
string evolution in more detail in Section IV. If a random tangle of
strings was formed in the early universe, there would always be some
strings longer than the horizon, so a few would remain even today.
Because cosmological phase transitions typically happen in the very
early universe, cosmic strings contain a lot of trapped energy, and
can therefore significantly perturb the matter distribution. To first
order there is a single parameter quantifying the effects of strings,
its energy per unit length. In the simple relativistic strings, the
mass per unit length and the string tension are equal, because of
Lorentz invariance under boosts along the direction of the string (but
this need not be true for more elaborate models, see
section VI. Cosmic strings are exceedingly thin, but very massive.
Typically, for strings produced around the epoch of grand
unification, the mass per unit length would be of order $\mu\sim
10^{21}$kg m$^{-1}$ and their thickness $10^{-24}$ m. The
gravitational effects of strings are effectively governed by the
dimensionless parameter $G\mu$, where $G$ is Newton's constant. For
GUT-scale strings, this is $10^{-6}$, while for electroweak-scale
strings it is $10^{-34}$. 

\section{String Formation}

Spontaneous symmetry breaking is a ubiquitous feature of our theories
of fundamental particle interactions. Cosmic strings are formed in
many symmetry-breaking phase transitions. If the symmetry is broken
from a group $G$ down to a subgroup $H$, the set of degenerate vacuum
or ground states is the manifold $M = G/H$, and the topology of this
manifold determines the types of defect that can form. In our previous
examples $M$ was a circle; in general, strings can form 
if $M$ contains closed curves that cannot be
contracted within $M$ (the technical term is that $M$ is not simply
connected, or that its first homotopy group is non-trivial) \cite{Kibble76}.

The Kibble mechanism described in section II
relates the initial density of strings to the size $\xi$ of the
domains over which the field is correlated,
\[ \rho_{\rm string} \sim C {1 \over \xi^2} \] with $C$ a constant of
order one reflecting the probability that a strings forms when three
or more domains meet. The correlation length cannot grow faster than
the speed of light so in the early universe an obvious upper bound on
$\xi$ is the size of the horizon at the time of the phase transition.
If the dynamics of the phase transition is known, $\xi$ can be
estimated more accurately. In a first order phase transition $\xi$ is
given by the typical distance between bubble nucleation sites, which
depends on the nucleation rate. In second order phase transitions
$\xi$ depends on the critical exponents and the rate of cooling
through the critical temperature, $T_c$, as shown by Zurek \cite{Zurek85,Zurek96}.

Vortex lines or topological strings can therefore appear in a wide
range of physical contexts, from cosmic strings in the early universe
through disclinations in room-temperature nematic liquid crystals, to
magnetic flux tubes in some superconductors and vortex lines in
low-temperature superfluid helium. These systems provide us
with a range of opportunities to test aspects of the cosmic string
formation and evolution scenario experimentally.

The Kibble mechanism in first order transitions was confirmed in
experiments on nematic liquid crystals \cite{Chuang91,Bowick94}. The Kibble-Zurek scenario for second order transitions has
been experimentally verified in Superfluid $^3$He \cite{Baeuerle96,Ruutu96} and in
Josephson Tunneling Junction arrays \cite{Monaco06}.

The He--3 experiments in a rotating cryostat in Helsinki also
confirmed the scale invariance of the initial distribution of loops,
$n(R) \sim R^{-4}$, where $n(R)dn$ is the number density of loops with
radii between $R$ and $R+dR$, as predicted by Vachaspati and Vilenkin \cite{Vachaspati84}.

More recently, the formation of a defect network following the
annihilation of $^3$He--A / $^3$He--B boundary layers has been
observed \cite{Bradley08}. The precise type of defects is still
under investigation but this system constitutes an interesting
analogue to the formation of strings from the annihilation of branes
in brane inflation scenarios.

There are also a few systems where the string density disagrees with
the Kibble-Zurek predictions. In $^4$He \cite{Dodd04}, the reasons are
understood: the strings are fuzzy and the network does not survive
long enough to be detected \cite{Karra98}. In the case of
superconducting films the results are somewhat inconclusive \cite{Maniv03} and also it is
not completely clear what the expected density of flux quanta should
be after a temperature quench; an alternative formation mechanism with
different vortex clustering properties has been proposed in \cite{Hindmarsh00}. In fact the formation of defects in
systems with gauge fields is clearly very relevant to cosmology but is
still not completely understood (see \cite{Kibble07} for a recent discussion).

\section{String Evolution}

The motion of a cosmic string with worldsheet coordinates $\sigma^a$
and background space-time coordinates $x^\mu$ in a metric $g_{\mu\nu}$
is obtainable from a variational principle applied to the Goto-Nambu
action \cite{Nambu70,Goto71}
\begin{equation}
S = \mu \times \textrm{Area} = 
\mu \int d\tau d\sigma | det g_{ab} |^2
= \mu \int d\tau d\sigma \biggl| det \pmatrix{
\dot{x^\rho}\dot{x^\nu} g_{\rho\nu}  &
\dot{x^\rho}{x^\nu}' g_{\rho\nu}  \cr
{x^\rho}'\dot{x^\nu}g_{\rho\nu}  &
{x^\rho}' {x^\nu}' g_{\rho\nu}  \cr} \biggr|^{1/2}
\end{equation}
where $\mu$ is again the string mass per unit length, and with dots
and primes respectively denoting derivatives with respect to the
time-like $(\tau)$ and space-like $(\sigma)$ coordinates on the
world-sheet. $g_{ab}$ is called the induced metric. We are interested
in strings in a FRW background space-time (see equation
(\ref{FRWtau})) and can choose worldsheet coordinates that make the
induced metric diagonal
\begin{equation}
\sigma^0=\tau\, , \qquad {\bf {\dot x}}\cdot{\bf x}'=0 \, ,
\label{newgauge2}
\end{equation}
The choice of conformal time coordinate simplifies the microscopic
evolution equations, although as we shall see later on physical time
is a more natural choice for the macroscopic evolution (see Section II
for definitions of the two time choices). It is also useful to define
the coordinate energy per unit $\sigma$,
\begin{equation}
\epsilon^2=\frac{{\bf x}'{}^2}{1-{\dot {\bf x}}^2}\, .\label{ceus}
\end{equation}
Then the usual variational techniques can be used to show that the microscopic
string equations of motion are
\begin{equation}
{\ddot {\bf x}}+2\frac{{\dot a}}{a}{\dot {\bf x}}(1-{\dot {\bf x}}^2)=
\frac{1}{\epsilon}\left(\frac{{\bf x}'}{\epsilon}\right)'\, \label{geqn1}
\end{equation}
and
\begin{equation}
{\dot\epsilon}+2\epsilon\frac{\dot a}{a}{\dot {\bf x}}^2=0\, .\label{gneq2}
\end{equation}
For simplicity we are neglecting effects such as cusps and a frictional 
force due to particle scattering (which
for heavy strings is only relevant during a transient period very
early in the network's evolution). The first one is just a
wave equation with a particular damping term (provided by the
expansion of the universe). The damping also has the effect of
reducing the coordinate energy per unit $\sigma$.

As was mentioned earlier the expansion of the universe stretches the
strings, so in the absence of energy loss mechanisms their energy
would grow with the scale factor and the string network would
eventually become the dominant component of the universe's energy
density---which would be in conflict with observational results. Such
decay mechanisms do exist (at least for the simplest models), being
ultimately due to radiation losses and to the fact that
whenever strings interact they will reconnect \cite{Shellard87,Moriarty88}. In particular
closed loops may be formed, and these subsequently oscillate and
eventually decay. This decay is thought to be mainly into gravitational
radiation, but other forms of radiation are also produced very
efficiently. 

Provided the decay rate is high enough, the network will not have
pathological consequences, but will instead reach a linear scaling
solution, where the string density is a constant fraction of the
background density and on large scales the network looks the same (in
a statistical sense) at all times. Scaling is in fact an attractor
solution, as has been shown both using numerical simulations and
analytic models. Physically, the reason for this is that if one has a
high density of strings then the number of string interactions
increases and therefore loop production becomes more efficient and the
decay rate increases. Conversely if the density is too low then there
are few interactions and the decay rate is correspondingly lower.
Numerical simulations confirm this broad picture, but also reveal that
string evolution is a complex non-linear process, involving
non-trivial interactions between various different scales.

There have been thus far two generations of numerical simulations of
Goto-Nambu cosmic string networks in expanding universes. The first
(Albrecht and Turok \cite{Albrecht89}, Bennett and Bouchet \cite{Bennett90},
Allen and Shellard \cite{Allen90}) dates from around 1990, at the peak of
the interest in cosmic strings as possible seeds for the large-scale
structures we observe today. In the last few years, the renewed
interest in strings in the context of models with extra dimensions led
to a second generation of simulations (Martins and Shellard \cite{Martins06},
Ringeval et al. \cite{Ringeval07}, Vanchurin et al. \cite{Vanchurin07}), which build upon previous knowledge and
exploit the dramatic improvements in hardware and software in the
intervening decade and a half to achieve a much higher resolution.

A different approach is provided by full field theory simulations \cite{Vincent98}. These are closer to the
microphysics of the defects and provide unique information on
the interactions of the defects and their energy loss mechanisms, but their
shorter dynamic range means that they are not optimal for understanding the 
non-linear feedback mechanisms between widely different scales which affect 
the dynamics of the network. From this point of view they play a very important
role as calibrators, both for Goto-Nambu simulations and for analytic models.
One can also carry out Minkowski space simulations (either of Goto-Nambu or 
field theory type). Neglecting the expansion of the universe is numerically
desirable, since such simulations are much easier to implement and
evolve much faster. However, the expansion plays a non-trivial role in
the network dynamics, so these results should not be naively
extrapolated to realistic cosmological scenarios.

Initial conditions for the numerical simulations are usually set up
using the Vachaspati-Vilenkin algorithm \cite{Vachaspati84}. One often adds to this random initial velocities, since
these tend to enhance the rate of relaxation. All simulations agree on the 
broad, large-scale features of string networks, and in particular on the fact
that the linear scaling solution is an attractor for the evolution.
In Goto-Nambu simulations, the initial fraction of the total
energy in the form of closed loops is around $20\% $, but in the
linear scaling regime this fraction is around $50\% $ or even slightly
more. On the other hand, in field theory simulations this fraction tends to
be somewhat smaller.

The first-generation simulations suggested a dynamical picture where
the long-string network lost energy to large, long-lived loops, with
sizes of order the correlation length. Refinements had each loop
self-intersecting into around 10 daughter loops, but loop production
from the long strings was essentially monochromatic. The
second-generation simulations, however, reveal a quite different
picture. Large loops do self-intersect (and indeed the number of
daughter loops produced by each one seems to around 20), but
there is also a direct production of large quantities of small loops
from slow-moving long-string with fractal-like substructure. In other
words, the loop production is in fact bi-modal. All three
second-generation simulations agree on this broad picture, though not
on which of the two loop production scales is dominant.

The second-generation simulations present some tentative evidence for the 
scaling of small-scale features of the network. An open question is whether
or not this is expected to happen, given that gravitational backreaction
(which would provide a characteristic scale) is not included in any of the 
network simulations carried out to date.
One possible explanation stems from the fact that large
loops are not scaled up versions of small loops. Indeed, small loops
tend to be nearly circular, whereas large loops are not only far from
circular but even far from planar. In other words, the
self-intersection probability for a given loop depends on its size,
and this may be sufficient to dynamically select a preferred
scale. Incidentally, the loop fragmentation processes in these
networks highlight the fact that there is a steady flow of energy
from large to small scales which is entirely analogous to a Richardson
cascade in turbulence. (In this case energy enters via long strings
falling inside the horizon, and leaves via radiative decays.)

\begin{figure}
\includegraphics[width=3.5in,keepaspectratio]{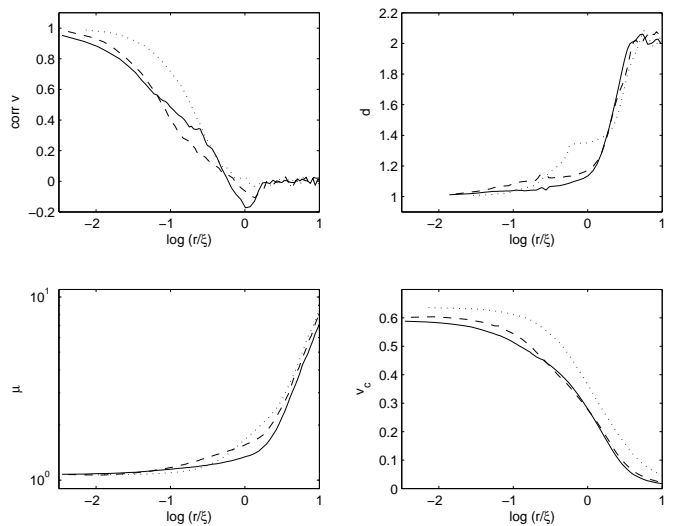}
\caption{\label{figfrac}Characteristic small-scale properties of
cosmic string networks in the linear scaling regime, for matter
(solid lines), radiation (dashed) and Minkowski spacetime (dotted)
runs. In all plots the horizontal axis represents the logarithm of
the physical lengthscale relative to the correlation length of the
network. The simulations leading to these results are described in \protect\cite{Martins06}.
Top panels show the correlation function
for the velocity vectors and the fractal dimension, bottom panels
show coarse-grained mass per unit length and coherent velocity.
The plotted quantities are described in the main text.}
\end{figure}

Figure \ref{figfrac} shows some relevant quantities characterizing
Goto-Nambu string networks in the linear scaling
regime. These are always plotted as a function of scale, relative to
the network's correlation length.  The top left panel shows the
correlation function for the velocity vectors. A first striking
feature is that in the expanding universe cosmic string velocities are
anti-correlated on scales around the correlation length (which is
smaller than but comparable to the causal horizon), but such a feature
is not present in Minkowski space. This anti-correlation is the result
of a `memory' of the network for recent reconnection events, and it is
ultimately due to the damping effect of expansion. The top right panel
depicts the fractal dimension of the network: this interpolates
between $d=1$ (straight segments) on small scales and $d=2$ (Brownian
network) on large scales, but it does so in a non-trivial way (which
is again different depending on whether or not there is expansion) and
over a wide range of scales. The fractal dimension evolves with time,
decaying on any given physical scale: the strings continually become
smoother on any scale, so as to minimize energy. Finally, the bottom
panels show the renormalized (or 'coarse-grained') string mass per
unit length on the left, and the corresponding coherent velocity on
the right panel---notice that the effect of expansion is to reduce the
velocities on any given scale.

The technical difficulty and computational cost of numerical
simulation provide strong motivation for alternative analytic
approaches, which essentially abandon the detailed statistical
physics of the string network to concentrate on its thermodynamics.
The best example is the velocity-dependent one-scale (VOS) model \cite{Martins02}, which
builds on previous work by Kibble and Bennett and has demonstrated
quantitative success when compared with both field theory and
Goto-Nambu numerical simulations. The 'one-scale' assumption is that
the network has a single characteristic lengthscale, which coincides
with the string correlation length and the string curvature
radius. This is an approximation which can be tested numerically.

The first assumption in this analysis is to localize the string so
that we can treat it as a one-dimensional line-like object. This is
clearly a good assumption for gauged strings, such as magnetic flux
lines, but may seem more questionable for strings possessing
long-range interactions, such as global strings or superfluid vortex
lines. However, good agreement between the VOS model and simulations
has been found in both cases. The second step is to average the
microscopic string equations of motion to derive the key evolution
equations for suitable macroscopic quantities, specifically its energy
$E$ and RMS velocity $v$ defined by
\be
E=\mu a(\tau)\int\epsilon d\sigma\, , \qquad v^2= \langle {\dot{\bf x}}^2
\rangle = \frac{\int{\dot{\bf x}}^2\epsilon
  d\sigma}{\int\epsilon d\sigma}\,. \label{eee}
\ee
Notice that the energy is an 'extensive' quantity but the RMS velocity is an
averaged quantity (and the averaging is weighted by the coordinate energy
$\epsilon$). In keeping with the above coordinate choices, the microscopic
quantities (on the right-hand side of both equations) are defined in terms of
conformal time, but it turns out that the macroscopic evolution that we are
now considering is best described in terms of physical time---please refer to
the cosmology review for the explicit relation between the two.

Any string network divides fairly neatly into two distinct
populations, \textit{viz.} long (or `infinite') strings and small
closed loops. In the following we will focus on the long strings.  The
long string network is a Brownian random walk on large scales and can
be characterised by a correlation length $L$, which can be used to
replace the energy $E= \rho V$ in long strings in our averaged
description, that is, \be
\rho \equiv {\mu \over L^2}\,.  \ee A phenomenological term
must then be included to account for the loss of energy from long
strings by the production of loops, which are much smaller than $L$. A
\textit{loop chopping efficiency} parameter $\tilde c$ is introduced
to characterise this loop production as \be
\left(\frac{d\rho}{dt}\right)_{\rm to\ loops}={\tilde
  c}v\frac{\rho}{L} \, . \label{rtl} \ee In this approximation, we
would expect the loop parameter $\tilde c$ to be a constant;
comparison with numerical simulations suggests ${\tilde c}\sim0.23$.

From the microscopic string equations of motion, one can then average
to derive the evolution equation for the correlation length $L$, \be
2\frac{dL}{dt}=2HL(1 + {v^2})+{\tilde c}v \, , \label{evl0} \ee where
$H$ is the Hubble parameter defined in eq. (\ref{defH}).
The first term in (\ref{evl0}) is due to the stretching of the network
by the Hubble expansion which is modulated by the redshifting of the
string velocity, while the second is the loop production term. One can
also derive an evolution equation for the long string velocity with
only a little more than Newton's second law \be
\frac{dv}{dt}=\left(1-{v^2}\right)\left(\frac{k(v)}{L}-2Hv\right)\, .
\label{evv0}
\ee The first term is the acceleration due to the curvature of the
strings and the second is the damping term from the Hubble expansion.
Note that strictly speaking it is the curvature radius $R$ which
should appear in the denominator of the first term. In the present
context we are identifying $R=L$. The function $k(v)$ is
the \textit{momentum parameter}, defined by \be
k(v)\equiv\frac{\langle(1-{\dot {\bf x}^2})({\dot {\bf x}}\cdot {\bf
    u})\rangle} {v(1-v^2)}\, ,
\label{klod}
\ee with ${\dot {\bf x}}$ the microscopic string velocity and ${\bf
  u}$ a unit vector parallel to the curvature radius vector. For most
relativistic regimes relevant to cosmic strings it is sufficient to
define it as follows: \be k_{\rm r}(v)
=\frac{2\sqrt{2}}{\pi}\;\frac{1-8v^6}{1+8v^6}\,,
\label{krel}
\ee while in the opposite case ($v\rightarrow0$), we have the
non-relativistic limit $k_0=2\sqrt{2}/\pi$.

Scale-invariant attractor solutions of the form $L\propto t$ (or
$L\propto H^{-1}$) together with $v=const.$, only appear to exist when
the scale factor is a power law of the form \be a(t)\propto t^\beta\,
, \qquad 0<\beta=const. <1\, \,.
\label{conda0}
\ee
This condition implies that
\be
L\propto t\propto H^{-1}\, ,
\label{props}
\ee with the proportionality factors dependent on the expansion rate
$\beta$. It is useful to introduce the following parameters to
describe the relative correlation length and densities, defining them
respectively as \be L=\gamma t\,, \qquad \zeta \equiv \gamma^{-2}=
\rho t^2/\mu\,.  \ee By looking for stable fixed points in the VOS
equations, we can express the actual scaling solutions in the
following implicit form: \be \gamma^2=\frac{k(k+{\tilde
    c})}{4\beta(1-\beta)}\, ,\qquad
v^2=\frac{k(1-\beta)}{\beta(k+{\tilde c})}\, ,
\label{scalsol}
\ee where $k$ is the constant value of $k(v)$ given by solving the
second (implicit) equation for the velocity. It is easy to verify
numerically that this solution is well-behaved and stable for all
realistic parameter values.

If the scale factor is not a power law, then simple scale-invariant
solutions like (\ref{scalsol}) do not exist. Physically this happens
because the network dynamics are unable to adapt rapidly enough to the
changes in the background cosmology. An example of this is the
transition between the radiation and matter-dominated eras. Indeed,
since this relaxation to a changing expansion rate is rather slow,
realistic cosmic string networks are strictly speaking \textit{never}
in scaling during the matter-dominated era. Another example is the
onset of dark energy domination around the present day. In this case,
the network is gradually slowed down by the accelerated expansion, and
asymptotically it becomes frozen in comoving coordinates. The
corresponding scaling laws for the correlation length and velocity are
$L\propto a$ and $v\propto a^{-1}$.  

Despite its success in
describing the large-scale features of string networks, the VOS model
has the shortcoming of not being able to account for the small-scale
features developing on the strings as the network evolves, as clearly
shown by numerical simulations. This small-scale structure is in the
form of wiggles and kinks, and can be phenomenologically characterized
by its fractal properties, as we have sketched above. As a first
analytic simplification, the string wiggles can be characterized
through a renormalized string mass per unit length that is larger than
the bare (Goto-Nambu) mass. This effectively corresponds to
considering a model with a non-trivial equation of state (the relation
between the string tension and the mass per unit length), which turns
out to be one among a larger class of models known as elastic string
models. This kind of description has interesting parallels with the
coarse-graining approaches that are typical of condensed matter.

A more radical approach is to explicitly abandon the one-scale
assumption. This is done in the three-scale model \cite{Austin93}, which distinguishes between the characteristic
lengthscale (which is simply a measure of the total string energy in a
given volume) and the persistence length (which is defined in terms of
the invariant length along the string and corresponds to the
correlation length or inter-string distance). Additionally there is a
third lengthscale which approximately describes a typical scale of the
small-scale wiggles. This kind of description is in principle highly
flexible, though this can be considered a blessing and a curse. The
downside is that one is forced to introduce a large number of (almost
free) phenomenological parameters over which one has limited control
even when comparing the model with simulations.

Having said that, the three-scale model does confirm, at least
qualitatively, the expectations for the behavior of string networks.
Scaling of the large scales (in this case the characteristic and
persistence lengths) is found to be an attractor, just as in the VOS
model. Depending on the behavior of small-scale structures, the two
large length scales may reach scaling simultaneously or the former may
do so before the latter---a behavior that has been seen in numerical
simulations. As for the behavior of the small-scale structures, their
evolution timescale is typically slower, and generically they only
reach scaling due to the effects of gravitational backreaction (not
included in numerical simulations). In the absence of gravitational
backreaction, scaling of the small-scale characteristic length is
contingent of the removal of a sufficiently large amount of
small-scale structure from the long strings by radiation
and loop production, which in the model is controlled by a parameter
whose detailed behavior is not known.

Finally, an interesting and rather different approach starts out with
the assumption that there is a range of scales where stretching due to
the expansion is the dominant dynamical effect, even on scales well
below the cosmological horizon. A sufficient condition for this is
that one is assuming that the rate of string intercommutations is
fixed in horizon units. This turns out to be sufficient to allow the
construction of a statistical-type description based on two-point
correlation functions \cite{Polchinski06}.
Their results are to a first approximation dependent on a critical
exponent which physically is related to the coherent string velocity
on a given scale. Comparison with numerical simulations shows, as
expected, that the best agreement is found around and just below the
horizon scale.

A second assumption is that loop production at those scales is
sufficiently localized to be describable as a  perturbation. When loop
production is thus folded into the analysis, the picture that
ultimately emerges is of a complicated fragmentation cascade. In
particular, this model provides supporting evidence for the
two-population loop distribution picture outlined above and clearly
seen in high-resolution simulations. There is a population of
correlation-length sized loops, produced by direct long-string
intercommutation, and a second population with sizes a few orders of
magnitude below (quite possibly near the gravitational backreaction
scale) and due to loop fragmentation. Whether or not the smoothing
provided by gravitational radiation is necessary to yield scaling of
the loop sizes is again not entirely clear at the moment, but it is in
principle a question for which this formalism could provide an answer.

\section{Astrophysical and Cosmological Consequences}

As was mentioned in Section I, the
spacetime around a straight cosmic string is flat. A string lying
along the $z$-direction has an equation of state $p_z = - \rho$,
$p_x = p_y = 0$ and therefore there is no source
term in the relativistic version of the Poisson equation 
for the Newtonian gravitational potential \be
\nabla^2\phi=4\pi G(\rho +p_x + p_y + p_z) = 0\,.  \ee A straight
string exhibits no analogue of the Newtonian pull of gravity on any
surrounding matter. However, this does not mean the string has no
gravitational impact at all. On the contrary, we will see that a
moving string has dramatic effects on nearby matter or propagating
microwave background photons. It is not difficult to derive the
spacetime metric about such a straight static string \cite{Vilenkin81}. It has the simple form \be ds^2 = dt^2 - dz^2 -
dr^2 - r^2d\theta^2\,,\ee which looks like Minkowski space in
cylindrical coordinates, except for the fact that the azimuthal
coordinate $\theta$ has a restricted range $0\le\theta \le
2\pi(1-4G\mu)$. That is, the spacetime is actually conical with a
global deficit angle \be \alpha=8\pi G\mu\,; \ee where an angular
wedge of width $\alpha$ is removed and the remaining edges identified.

\begin{figure}
\includegraphics[width=3.5in,keepaspectratio]{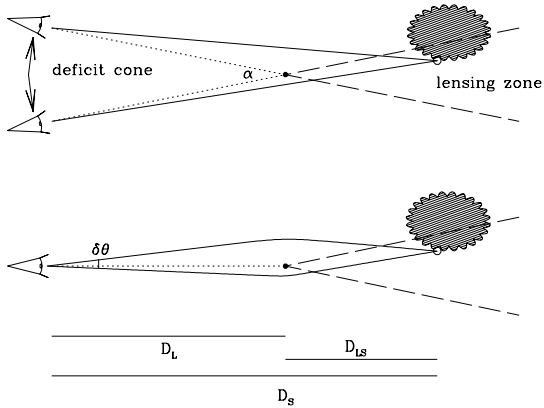}
\caption{\label{figlens}An illustration of the mechanism behind lensing by
cosmic strings. The thick black dot represents a cosmic string perpendicular
to the page. The spacetime metric around the string can be obtained by
removing the angular wedge of width $\alpha$ and identifying the edges.
An observer can thus see double images of objects located on a certain zone
behind the string. This zone is enclosed by the dashed lines, while the solid
lines depict light rays and the angular separation of the two images,
$\delta\theta$, will depend on the distances of the source and the observer to
the string as well as on the deficit angle. Reprinted, with permission, from \protect\cite{Kuijken07}.}
\end{figure}

This deficit angle implies that the string acts as a cylindrical
gravitational lens, creating double images of sources behind the string (such
as distant galaxies), with a typical angular separation $\delta\theta$ of
order $\alpha$ and no distortion \cite{Vilenkin84}. This is illustrated in
Figure \ref{figlens}. A long string would yield a distinctive lensing pattern.
We should expect to see an approximately linear array of lensed pairs, each
separated in the transverse direction. In each lensing event the two images
would be identical and have essentially the same magnitude. (Except if we
happen to see only part of one of the images.) This is a very unusual
signature, because most ordinary gravitational lenses produce an odd
number of images of substantially different magnitudes. A number of
string lensing event candidates have been discussed in the past, but
no confirmed one is currently known.

However, the above simple picture is complicated in practice by the
fact that cosmic strings are not generally either straight or static.
Whenever strings exchange partners kinks are created that straighten
out only very slowly, so we expect a lot of small-scale structure on
the strings. Viewed from a large scale, the effective tension and
energy per unit length will no longer be equal. Since the total length
of a wiggly string between two points is greater, it will have a
larger effective energy per unit length, $U$, while the effective
tension $T$, the average longitudinal component of the tension force,
is reduced, so $T < \mu < U$. This means that there is a non-zero
gravitational acceleration towards the string, proportional to $U-T$.
Moreover, the strings acquire large velocities, generally a
significant fraction of the speed of light, which introduces further
corrections to the deficit angle.

Another effect is the formation of over-dense wakes behind a moving
cosmic string \cite{Silk84}. When a string passes between
two objects, these are accelerated towards each other to a velocity
\be u_\perp = 4\pi G\mu v\,, \ee where $v$ is the string velocity.
Matter therefore collides in a sheet-like structure, leaving a wake
behind the moving string. This was the basic mechanism underlying the
formation of large-scale structures in cosmic string models. This
model has significant attractions, such as the early formation of
nonlinear structures, and one can get a good match to the observed
galaxy power spectrum in models with a large cosmological constant.
However, as we shall discuss, it fails to reproduce the power spectrum
of CMB anisotropies observed by COBE, WMAP and other experiments;
cosmic strings, therefore, can only play a subdominant role in
structure formation (albeit still significant, at the ten to twenty
percent level).  Cosmic strings create line-like discontinuities in
the cosmic microwave background signal \cite{Kaiser84,Gott85}. For the same reason that wakes form behind a cosmic
string, the CMB source on the surface of last scattering is boosted
towards the observer, so there is a relative CMB temperature shift
across a moving string (a red-shift of the radiation ahead of it, and
a blue-shift of that behind), given by \be \frac{\delta T}{T}\sim8\pi
G\mu v_\perp\,.  \ee where $v_\perp$ is the component of the string
velocity normal to the plane containing the string and the line of
sight. This is known as the Kaiser-Stebbins effect. This simple
picture is again complicated in an expanding universe with a wiggly
string network and relativistic matter and radiation components. The
energy-momentum tensor of the string acts as a source for the metric
fluctuations, which in turn create the temperature anisotropies. The
problem can be recast using Green's (or transfer) functions which
project forward the contributions of strings at early times to today \cite{Durrer01}.
The actual quantitative solution of this problem entails a
sophisticated formalism to solve the Boltzmann equation and then to
follow photon propagation along the observer's line of sight. At the
time of writing, the most recent comparisons \cite{Bevis07} between full-sky maps of cosmic
string-induced anisotropies and WMAP data yield a cosmological
constraint on the models with \be {G\mu} < {\rm few}\times 10^{-7}\,,
\ee with only a weak dependence on the background cosmology---in
particular, on the magnitude of the cosmological constant.

\begin{figure} 
  \vspace{-2.55mm}
  \resizebox{\columnwidth}{!}{\includegraphics{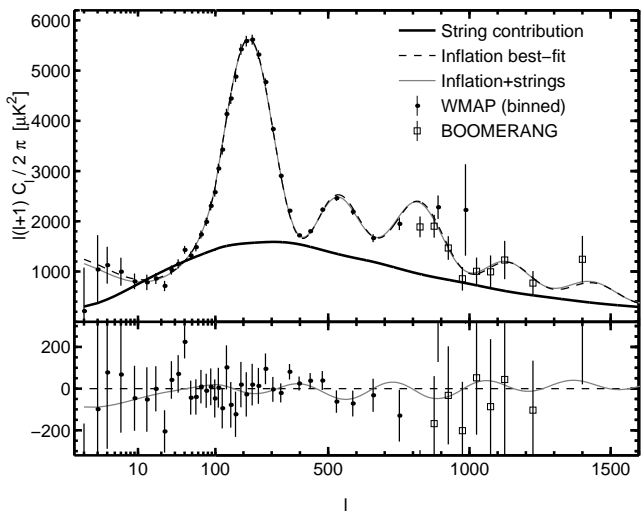}}
  \caption{\label{TT} The CMB temperature power spectrum contribution
    from cosmic strings, normalized to match the WMAP data at
    \mbox{$\ell=10$}, as well as the best-fit cases from inflation
    only (model PL) and inflation plus strings (PL+S). These are
    compared to the WMAP and BOOMERANG data. The lower plot is a
    repeat but with the best-fit inflation case subtracted,
    highlighting the deviations between the predictions and the data.
    Reprinted, with permission, from \protect\cite{Bevis07}.}
\end{figure}

Apart from their scale-invariance (which follows from the network's
attractor scaling solution discussed in the previous section), cosmic
defect-induced fluctuations appear to be the antithesis of the
standard inflation paradigm, because they are causal or active (they
are generated inside the horizon and over an extended period), there
are also large vector and tensor contributions, and they are
distinctly non-Gaussian.  All these characteristics leave clear
signatures in the cosmic microwave background angular power spectrum,
chief of which is a much broader primary Doppler peak and little
evidence of secondary oscillations. Unlike inflation, active defect
sources act incoherently with extra large-scale power from the vectors
and tensors. Moreover, their isocurvature nature provides a partial
explanation for why the broad primary peak ends up at larger
multipoles (typically $\ell\sim 300$ as opposed to $\ell\sim 200$ in a
flat cosmology. Isocurvature perturbations correspond to equal and
opposite perturbations in the radiation and matter densities --as
opposed to equal fractional perturbations in the number densities of
the two components for adiabatic perturbations--.  On the basis of
knowledge from present simulations, therefore, cosmic defects alone
are extremely unlikely to have been the seeds for large-scale
structure formation. However, they cannot be ruled out entirely. For
example, admixtures of inflationary power spectra with significant
cosmic defect contributions (at a level around $20\% $, see Figure
\ref{TT}) do provide a satisfactory fit to present data. This is
interesting among other reasons because it is the sort of level at
which the non-Gaussian signatures of cosmic strings should still be
discernible, although their distinct line-like discontinuities are
only clearly identifiable on small angular scales around a few arc
minutes.

Accelerated cosmic strings are sources of gravitational radiation \cite{Vilenkin81b}.
Consequently, a network of long strings and closed
loops produces a stochastic gravitational wave background \cite{Hogan84}
over a wide range of frequencies and with a spectrum which
(at least to a first approximation) has equal power on all logarithmic
frequency bins. Another distinctive signal would come
from the cusps, the points at which the string instantaneously doubles
back on itself, approaching the speed of light. Such an event
generates an intense pulse of gravitational and other types
of radiation, strongly beamed in the direction of motion of the cusp \cite{Damour00}. If massive cosmic strings do
indeed exist, both these pulses and the stochastic background are
likely to be among the most prominent signals seen by the
gravitational-wave detectors now in operation or planned, in
particular LIGO and LISA.

A stringent, though indirect, limit on the string
energy per unit length comes from observations of the timing of
millisecond pulsars. Gravitational waves between us and a pulsar would
distort the intervening space-time, and so cause random fluctuations
in the pulsar timing. The fact that pulsar timing is extremely regular
places an upper limit on the energy density in gravitational waves,
and hence on the string scale. The upper limit \cite{Jenet06} is of order $G\mu<10^{-7}$,
though there is still considerable uncertainty because
this depends on assumptions about the evolution of small-scale structure.

Although gravitational waves are thought to be the main decay byproduct of the
evolution of the simplest cosmic string networks, direct decay into
particle radiation is extremely efficient and there are claims that it
could be the dominant energy-loss mechanism responsible for scaling \cite{Vincent98}.
In more complicated models, there are certainly other decay
channels. If the strings are global (rather than local), then they
will preferentially produce Goldstone bosons instead. In axion models,
these Goldstone bosons acquire a small mass and become the axions (a
prime dark matter candidate). One can estimate the number density of axions
using analytic models of cosmic string evolution such as the VOS
model. In another class of string models known as superconducting
(since they have additional current-like degrees of freedom), then the
decay products can include electromagnetic radiation.

Finally, we should mention the claims that cosmic strings could be
responsible for a number of high energy astrophysical and cosmological
enigmas, including ultra-high energy cosmic rays, gamma ray bursts,
and baryogenesis (the creation of the matter-antimatter asymmetry of
the universe). Since cosmic defects can produce high energy particles,
they could contribute to the observed cosmic ray spectrum, notably at
ultra-high energies $E\ge10^{11}GeV$ where the usual acceleration
mechanisms seem inadequate \cite{Hill87}. Many ideas have been explored, such as particle emission
from cosmic string cusps, but most have been found to produce a
particle flux well below current observational limits. However, among
the interesting scenarios deserving further study are those with
hybrid defects (such as monopoles connected by strings) or vortons.

\section{Field Theory Strings with More Degrees of Freedom}

Kibble's original idea was to consider strings in grand unification
scenarios, in which the strong and electroweak forces become unified
at an energy scale of around $10^{15-16}$GeV. More recent studies have
shown that practically any viable Supersymmetric Grand Unified Theory
has a pattern of symmetry breaking transitions that leads to the
possibility of cosmic string formation at some point in its history \cite{Jeannerot03}.  In
particular, in these models the inflationary phase generically ends
with a phase transition at which strings are produced. More recent
studies suggest that this is also a feature of brane inflation models.

So far we have discussed the field theory realization of the simplest
model of cosmic string, the Abrikosov--Nielsen--Olesen string, in
which the mass per unit length equals the tension and there is no
internal structure apart from the magnetic field. In fact, the
situation can be much more complex in the early Universe, and
realistic particle physics models lead to networks with a much richer
phenomenology.  The added complexity makes these networks
much harder to study, whether by analytic or numerical methods, and
consequently they are not as well understood as the simplest case.

We can only give here a brief description of the possible
complications. The list below is not complete and furthermore there
are strings that fit more than one category. For a more detailed
discussion we refer the reader to the reviews by Vilenkin \& Shellard \cite{Vilenkin},
Hindmarsh \& Kibble \cite{Hindmarsh}, Carter \cite{Carter}, and Ach\'ucarro and Vachaspati \cite{Achucarro}, where
references to the original literature can also be found.

\subsection{Wiggly strings; varying tension string networks; cycloops}.

The name wiggly strings is sometimes used to refer to any type of
string whose mass per unit length is different from its tension. We
already mentioned in Section V that small structure (wiggles) on
the string produces a renormalized effective mass per unit length
$U>\mu$ and an effective tension $T < \mu$. There are other effects
that can affect the mass and tension, for instance the presence of
currents along the strings.

A particular kind of small structure is found in extra dimensional
models. If spacetime has more than the three spatial dimensions we
observe, strings may be able to wrap around the extra dimensions in
different ways leading to a renormalized four--dimensional tension and
mass per unit length. The effective tension can of course vary along
the strings. In extreme cases, the extra dimensional wrapping effects
concentrate around certain points along the string which behave like
'beads' (see hybrid networks below) and are called cycloops.

As discussed towards the end of Section IV,
the additional degrees of freedom (which can be thought of as a mass
current) make the evolution of the networks highly non-trivial. The
one-scale assumption is no longer justified: the correlation length,
inter-string distance and string curvature radius become distinct
lengthscales. Depending on the exact interplay between the bare
strings and the mass current wiggles, these lengthscales
can evolve differently, and some of them might be scaling while the
others are not. The presence of extra dimensions provides a further
energy flux mechanism (as energy may be lost into or gained from the
extra dimensions) which will affect the string dynamics, but at the
time of writing its exact effects have not been studied in detail.

\subsection{Non--topological / embedded / electroweak / semilocal strings}

In the Abrikosov--Nielsen--Olesen case, the scalar field is zero at
the core of the string, and the symmetry is unbroken there. The zero
field is protected by the topological properties of the vacuum
manifold (the non-contractible circle) and the string is called
topological.  In more realistic models, the criterion for topological
string production is a non-simply connected vacuum manifold, however
complicated. These strings are unbreakable and stable.

On the other hand there are examples in which there is no topological
protection but the strings are nevertheless stable. The scalar field
configuration at the core can be deformed continuously into a ground
state, so these non-topological strings can break, their magnetic flux
can spread out, or be converted to a different type of flux. But
whether this happens is a dynamical question that depends on the
detailed masses and couplings of the particles present, on the
temperature, etc.

The best studied examples of non--topological strings look like
Abrikosov--Nielsen--Olesen strings 'embedded' in a larger model such
as the Glashow--Salam--Weinberg model of electroweak interactions \cite{Nambu77,Vachaspati92}.
These \emph{electroweak} strings carry magnetic flux
of the Z boson, but the strings would only be stable for unphysical
values of the Z boson mass. They are closely related to
\emph{semilocal} strings, another example of embedded strings where the
symmetry breaking involves both local and global symmetries
intertwined in a particular way. For low scalar mass these can be
remarkably stable.

In general, non--topological strings are not resilient enough for the
networks to survive cosmological evolution. If the strings are
unstable to spreading their magnetic flux, the network will not form.
If the strings are breakable the network may form initially but it
will quickly disappear (see hybrid networks below for a concrete
example).  A remarkable exception to this rule are semilocal strings
with very low scalar mass: the network forms as a collection of
segments which then grow and reconnect to form longer strings or
loops. These evolve like a network of Abrikosov--Nielsen--Olesen
strings plus a small population of segments and there is some evidence
of scaling \cite{Achucarro07}.

\subsection{Dressed / superconducting strings / vortons} 

In realistic particle physics models, a stable string will trap in its
core any particles or excitations whose mass is lower inside due to
the interactions with the scalar field. These \emph{dressed} strings
have a more complicated core structure. In extreme cases, the mass of
these trapped particles is zero in the core and they lead to
persistent currents along the strings, which are then known as
\emph{superconducting} \cite{Witten85}. In some cases, the decay of a loop of
superconducting string can be stopped by these currents, leading to
long-lived remnants called \emph{vortons} \cite{Davis89} that destroy scaling;
typical vortons will either dominate the energy density of the
universe (contrary to observations) or contribute to the dark matter
if they are sufficiently light.

\subsection{Hybrid networks}

Hybrid networks contain more than one type of defect, such as for
instance strings of different kinds or composite defects combining
strings, monopoles and/or domain walls. 

\subsubsection{Composite defects:}

The production of strings may be accompanied by the production of
other defects such as monopoles or domain walls, before or afterwards,
that change the behaviour of the network as a whole.  These networks
can have radically different scaling properties ---in particular,
linear scaling may not exist at all. Consider for instance a sequence
of breakings of the form $G \to H \to K$ in which a symmetry group $G$
first breaks down to a subgroup $H$ which subsequently breaks to an
even smaller subgroup $K$ at a lower temperature. Two cases are
particularly relevant for strings:

The first breaking produces stable magnetic monopoles, the second
  confines --totally or partly-- the magnetic field to flux tubes
  (strings) leading to a network of monopoles connected by
  strings. This can happen either as string segments, with monopoles
  at the ends, which eventually contract and disappear or as a network
  of strings carrying heavy ``beads'' (the monopoles) which can lead
  to a scaling solution.

The first breaking produces stable strings, the second makes domain
  walls attached to the strings (e.g. in \emph{axion}
  models). The network is made of pancake-like structures that
contract under the wall tension and eventually disappear, although in
some cases there may be long-lived remnants.

\subsubsection{Non--abelian / (p,q) strings}

Another type of hybrid network contains different types of strings
whose intercommutation leads to three-point junctions and bridges.
These networks are also very different from the simplest ones but the
current consensus is that they also seem to reach a scaling solution
during cosmological evolution.  

In the non-abelian case, the magnetic flux carried by the string is
not just a number but can have different internal ``orientations''.
These become relevant when the strings cross, limiting the ways in
which they can reconnect.

Hybrid networks containing several interacting string types are also
found in superstring models (see next section).  The most interesting
type, usually referred to as $(p,q)$ strings, contains two types of
string each carrying different type of flux that is separately
conserved: fundamental and solitonic or D-strings, roughly
corresponding to electric and magnetic flux tubes.  The numbers $p$
and $q$ refer to the units of each kind of flux carried by the
strings.  Since the mass per unit length depends on these fluxes,
$(p,q)$ networks are expected to have a hierarchy of different
tensions, as well as junctions and bridges. In fact, junctions and
bridges will also form in any model in which parallel strings have an
attractive interaction, such as Abrikosov-Nielsen-Olesen strings with
extremely low scalar to vector mass ratios.

The existence of string junctions and the hierarchy of string tensions
make the evolution of these networks considerably more complicated
than that of the simple Goto-Nambu strings. Relatively simple analyses
suggest that the heavier strings with gradually decay into the lighter
ones, and scaling is eventually reached for the strings at the low end
of the spectrum (the heavier ones eventually disappear), although this
is still under discussion. Naive expectations that the network might
be slowed down to non-relativistic speeds and eventually freeze have
so far not been supported by the (admittedly simplistic) numerical
simulations performed so far. Further work is needed to understand the
general conditions under which scaling is (or is not) an attractor.

\section{Cosmic Superstrings}

Superstring theory is to date the only candidate model for a
consistent quantum theory of gravity that includes all other known
interactions.  In string theory, the fundamental constituents of
nature are not point-like particles but one-dimensional ``strings''
whose vibrational modes produce all elementary particles and
their interactions. Two important features of the theory are supersymmetry
(a symmetry between bosonic and fermionic excitations that keeps
quantum effects under control) and the presence of extra dimensions
above the four spacetime dimensions that we observe.

It is not yet known how to formulate the theory in its full generality
but some weak-coupling regimes are well understood. In these, the
fundamental strings live in a 10-dimensional spacetime, of which 6
dimensions are ``compactified'', resulting in an effective
4-dimensional spacetime we live in. There is another regime, M-theory,
in which the fundamental objects are two-dimensional ``membranes'' and
the background spacetime is 11-dimensional. These regimes are related
to one another by duality transformations that interchange the role of
fluctuation quanta and non-perturbative, soliton-like states (branes),
so the expectation is that all regimes are different limits of a
unique, underlying theory usually referred to as superstring/M-theory,
or just M-theory for short.

Before the discovery of D-branes, the ``solitons'' of superstring
theory, the question of whether fundamental superstrings could ever
reach cosmological sizes was analysed and the possibility discarded \cite{Witten85b}. There
were two main problems.  First, the natural mass per unit length of
fundamental strings is close to the Planck scale and would correspond
to deficit angles of order {\bf $2 \pi$}, which would have been
observed. Second, the strings were inherently unstable to either
breaking or --depending on the type of string-- becoming the boundary
of domain walls that would quickly
contract and disappear.  The discovery of branes and their role in
more exotic compactifications where the six compact dimensions have
strong gravitational potentials (and redshifts) have changed this
picture. It is now believed that networks of cosmic superstrings
could be a natural outcome of brane-antibrane annihilation,
especially if the branes are responsible for a period of cosmic
inflation \cite{Sarangi02,Jones03,Copeland04,Polchinski04}.  

An important difference with previous scenarios is that these strings
are located in regions of the compactified dimensions with very strong
gravitational redshift effects (``warping'') that reduce the effective
mass per unit length of the strings to a level with deficit angles in
the region of $10^{-12}$ to $10^{-7}$, compatible with current
observations.  Another important difference is a much lower
probability that the strings intercommute when they cross, estimated
to be $10^{-3}$ to $ 10^{-1}$, depending on the type of strings. The
lower intercommutation rates lead to much denser networks. Estimates
of the corresponding enhancement in the emission of gravitational
radiation by cusps puts these strings in a potentially observable
window by future gravitational wave detectors \cite{Damour05,Siemens07}.

The networks are hybrid, consisting of fundamental strings and
D-strings, the latter being either one-dimensional D-branes or perhaps
the result of a higher dimensional D-brane where all but one dimension
are wrapped around some ``holes'' (cycles) in the compactified space.
There may also be cycloops.

As in the case of hybrid field theory strings, whether or not
superstring networks eventually reach a scaling regime is an open
question. Analytic studies and numerical simulations of simplified
cases suggest that scaling is certainly possible, though contingent
on model parameters that at the time of writing are not well
understood. In this case, in addition to the presence of junctions
and a non-trivial spectrum of string tensions, a third factor can
affect to the evolution of these networks. If the strings are
actually higher-dimensional branes partially wrapped around some
extra dimensions, then energy and momentum can in principle leak
into or out of these extra dimensions \cite{Avgoustidis05}.  Since the effective damping force affecting the
ordinary and extra dimensions is different, one might generically
expect that this will be the case. Depending on its sign and
magnitude, such an energy flow can in principle prevent scaling,
either by freezing the network (if too much energy leaks out) or by
making the strings dominate the universe's energy density (if
too much energy leaks in, though this is less likely than the
opposite case). In this sense, a somewhat delicate balance may be
needed to ensure scaling. At a phenomenological level, further work
will be required in order to understand the precise conditions under
which each of these scenarios occurs. At a more fundamental level,
it is quite likely that which of the scenarios is realized will
depend on the underlying compactifications and/or brane inflation
models, and that may eventually be used as a discriminating test
between string theory realizations.

\section{Future Directions}

One of the most exciting prospects is the discovery of magnetic-type
CMB polarization (usually referred to as B-modes) as this would reveal
the presence of vector and/or tensor modes. Cosmic string models may
be further constrained in the near future because B-modes are
predicted to have amplitudes comparable to the electric-type E-modes
(at large angular scales). At high resolution, one
could also hope to observe defects directly through the B-mode signal,
against a relatively unperturbed background. Conversely, the detection
of vector modes would provide strong evidence against inflation
without cosmic defects. Polarization data will also strongly
constrain a significant isocurvature contribution to the mainly
adiabatic density fluctuations. Isocurvature perturbations 
can be a signature of more complicated physics
during inflation, such as the effects of two or more scalar fields, or
the formation of defects at the end of inflation.

Ongoing and future CMB experiments, especially at high resolution,
will be probing the degree of Gaussianity of the primordial
fluctuations. The detection of significant and unambiguous
non-Gaussianity in the primary CMB signal would be inconsistent with
simple (so called single field slow-roll) inflation. More general
inflationary models can accommodate certain types of non-Gaussianity,
and one can also envisage non-Gaussianity from excited initial states
for inflation. It is interesting to note that given the existing
bounds on $G\mu$, current CMB experiments do not have the sensitivity
or resolution to detect cosmic string signatures directly, in
particular the Kaiser-Stebbins effect in CMB maps. However, with
high-resolution sensitivities becoming available in the near future,
direct constraints (or detections) will be possible. This is just one
example of the interesting new science that future high-resolution CMB
experiments might uncover in the years ahead. In particular, ESA'a
Planck Surveyor [See www.rssd.esa.int/Planck/], scheduled for launch in
late 2008, may be able to provide significant breakthroughs.

A deeper understanding of the evolution and consequences of string
networks will require progress on both numerical simulations and
analytic modellings. At the time of writing there is still no
numerical code that includes all the relevant physics, even for the
simplest (Goto-Nambu) strings. Inclusion of gravitational backreaction
is particularly subtle, and may require completely new approaches. The
expected improvements in the available hardware and software will
allow for simulations with much longer evolution timespan and spatial
resolution, which are needed in order to understand the non-linear
interactions between large and small scales all the way down to the
level of the constituent quantum fields. This in turn will be a
valuable input for more detailed analytic modelling, that must
accurately describe the non-trivial small-scale properties of the
string networks as well as the detailed features of the loop
populations. Better modelling is also needed to describe more general
networks---three crucial mechanisms for which at present there is only
a fairly simplistic description are the presence of junctions, a
non-trivial spectrum of string tensions, and the flow of
energy-momentum into extra dimensions. 

At a more fundamental level, a better understanding
of the energy loss mechanisms and their roles in the evolution of the
networks is still missing \cite{Borsanyi07}
and it will require new developments in the theory of quantum fields
out of equilibrium.  Such theoretical developments are also needed to
understand defect formation in systems with gauge fields, and could be
tested experimentally in superconductors. 

The early universe is a unique laboratory, where the fundamental
building blocks of nature can be probed under the most extreme
conditions, that would otherwise be beyond the reach of any
human-made laboratory. Cosmic strings are particularly interesting for
this endeavour: they are effectively living fossils of earlier
cosmological phases, where physical conditions may have been
completely different. The serendipitous discovery of cosmic defects or
other exotic phenomena in forthcoming cosmological surveys would have
profound implications for our understanding of cosmological evolution
and of the physical processes that drove it. The search continues
while, in the meantime, the absence of cosmic string signatures will
remain a powerful theoretical tool to discriminate between fundamental
theories. The possibility that something as fundamental as superstring
theory may one day be validated in the sky, using tools as mundane as
spectroscopy or photometry, is an opportunity than neither
astrophysicists nor particle physicists can afford to miss.

\begin{acknowledgments}
A.A.'s work is supported by the Netherlands Organization for Scientific Research (N.W.O) under the VICI programme, and by the spanish government through the Consolider-Ingenio 2010 Programme CPAN (CSD2007-00042) and project  FPA 2005-04823.  The work of C.M.  is funded by a Ci\^encia2007 Research Contract.  
\end{acknowledgments}

\bibliography{review}
\end{document}